\documentclass[journal]{IEEEtran}
\IEEEoverridecommandlockouts

\usepackage{amssymb}
\usepackage[cmex10]{amsmath}
\usepackage{stfloats}
\usepackage{graphicx}
\usepackage{subfigure}
\usepackage{tabularx}
\usepackage{verbatim}
\usepackage{url}
\usepackage{bm}
\usepackage{booktabs}
\usepackage[colorlinks]{hyperref}

\usepackage{amsmath}
\usepackage{algorithm}
\usepackage{algorithmic}
\allowdisplaybreaks[2]
\usepackage{color}
\definecolor{myc1}{rgb}{0,0,0}
\usepackage{setspace}
\begin{document}

\title{Federated Unlearning Over Wireless Networks }

\author{Yixuan Chen, 
            Zhouxiang Zhao, 
            Wei Xu, \IEEEmembership{Fellow, IEEE,}
            Zhaoyang Zhang, \IEEEmembership{Senior Member, IEEE,}\\
           and Zhaohui Yang
            
 \vspace{-1.5em}
\thanks{Y. Chen, Z. Zhao, Z. Zhang, and Z. Yang are with the College of Information Science and Electronic Engineering, Zhejiang University, Hangzhou 310027, China  (e-mails: \{chen\_yixuan, zhouxiangzhao, ning\_ming, yang\_zhaohui\}@zju.edu.cn).}
\thanks{Wei Xu is with the National Mobile Communications Research Laboratory, Southeast University, Nanjing 210096, China, and also with the Purple Mountain Laboratories, Nanjing 211111, China (e-mail: wxu@seu.edu.cn).}

}

\maketitle

\begin{abstract}
To comply with stringent data privacy regulations, federated unlearning (FU) has emerged as a critical paradigm. However, its implementation over wireless networks introduces severe communication latency and reliability challenges due to iterative calibration requirements and physical-layer channel uncertainties. In this paper, we investigate the problem of delay minimization for federated unlearning networks (FUN). Specifically, we establish a comprehensive system model that jointly incorporates the convergence behavior of the FUN algorithm, local device computation dynamics, and a worst-case robust transmission model operating under bounded channel state information (CSI) error. To solve the resulting non-convex joint resource allocation problem, we propose an efficient iterative algorithm. By exploiting the monotonicity and convexity properties of the system constraints, the problem is decomposed via a uniform scan over the local accuracy parameter, within which the optimal delay, bandwidth, power, and computation frequency are determined utilizing nested bisection and golden-section searches. Both theoretical analysis and extensive numerical results demonstrate that the proposed algorithm achieves polynomial complexity and significantly reduces the overall unlearning completion time compared to conventional baseline schemes.
\end{abstract}

\begin{IEEEkeywords}
Federated unlearning, delay minimization, resource allocation, wireless communication, edge computing.
\end{IEEEkeywords}
\IEEEpeerreviewmaketitle

\section{Introduction}
\IEEEPARstart{T}{he} proliferation of edge devices, ranging from smartphones and wearables to Internet of things (IoT) sensors, has catalyzed an unprecedented explosion of data at the network edge. Traditionally, training artificial intelligence (AI) models required centralizing this data, incurring significant communication overhead and posing severe privacy risks \cite{1, 2, 3}. To mitigate these challenges, federated learning (FL) has emerged as a transformative paradigm, enabling distributed clients to collaboratively train a global model under the coordination of a central server without sharing raw local data \cite{4, 5, 6}. This decentralized framework has been successfully deployed in latency-sensitive and privacy-critical applications, including mobile keyboard prediction \cite{7}, healthcare analytics \cite{8, 9}, and intelligent transportation systems \cite{10, 11}.

Despite its privacy-preserving nature, standard FL protocols do not inherently support the right to erasure mandated by stringent data protection regulations such as the General Data Protection Regulation (GDPR) in Europe and the California Consumer Privacy Act (CCPA) in the United States \cite{12, 13, 14}. These regulations grant individuals the right to request the erasure of their personal data and its influence on trained models. Simply removing raw data from local storage is insufficient, as the information remains embedded in the global model parameters \cite{15, 16}. This necessity has given rise to the field of machine unlearning (MU) \cite{17, 18, 19}, which aims to efficiently eliminate the contribution of specific data samples or clients from a trained model. Extending MU to distributed settings leads to federated unlearning (FU) \cite{20, 21, 22}, a critical capability for maintaining trust and regulatory compliance in FL ecosystems.

Existing FU approaches can be broadly categorized into exact unlearning and approximate unlearning. Exact unlearning methods, such as retraining from scratch or frameworks, guarantee that the unlearned model is statistically identical to a model trained without the target data \cite{23, 24, 25}. However, these methods are often computationally prohibitive for large-scale deep learning models. Consequently, research has shifted towards approximate unlearning, which leverages techniques such as gradient ascent \cite{26, 27}, knowledge distillation \cite{28, 29, 30}, and influence functions \cite{31, 32} to approximate the retraining process with lower complexity. Notable contributions include FedEraser, which utilizes historical updates to reverse the learning process \cite{33}, and methods based on Fisher information matrix (FIM) for precise contribution removal \cite{34, 35}. More recent works have explored momentum degradation \cite{36} and parameter adjustment strategies \cite{37, 38} to enhance unlearning efficiency.

However, implementing FU in wireless edge networks introduces unique and stringent challenges that have not been fully addressed by existing algorithm-centric works \cite{39, 40}. Unlike standard FL, FU often requires iterative calibration rounds between the server and remaining clients to neutralize the influence of the leaving client \cite{41, 42}. This process significantly amplifies communication overhead, creating a bottleneck in bandwidth-constrained wireless environments. Furthermore, edge devices are typically energy-constrained, making it difficult to execute computationally intensive unlearning tasks locally \cite{43, 44}. A critical yet often overlooked factor is the impact of wireless channel dynamics. Practical systems suffer from channel fading, interference, and imperfect channel state information (CSI) due to estimation errors and feedback delays \cite{45, 46, 47}. These uncertainties can severely degrade the reliability of model update transmissions during the critical unlearning phase, potentially leading to incomplete unlearning or excessive latency.

Although recent studies have explored communication-efficient FL designs, such as client selection \cite{48,49} and update compression \cite{50,51}, the intersection of communication efficiency and FU has recently attracted increasing attention. For instance, the FATS framework was proposed to achieve exact FU with provable communication efficiency by leveraging total variation stability and periodic averaging \cite{52}. Similarly, compressed particle-based Bayesian protocols demonstrated that quantization and sparsification can reduce per-iteration communication overhead in FU while preserving Bayesian calibration benefits \cite{53}. Client selection strategies have also been tailored for FU, integrating deep reinforcement learning to dynamically select remaining clients for unlearning, thereby mitigating bias from unbalanced data distributions\cite{54}. Furthermore, hierarchical federated unlearning (Hier-FUN) was developed to address non-IID data and heterogeneous edge computing environments by organizing devices into tiers and confining most deletion traffic to lower tiers \cite{55}. Despite these algorithm-centric advances that primarily focus on reducing the number of communication rounds, few have explicitly optimized physical-layer resources under realistic channel uncertainties. Most existing works assume ideal communication links or focus solely on reducing the number of communication rounds, without jointly allocating bandwidth, transmit power, and computation frequency.

To bridge this gap, this paper investigates the problem of delay minimization for federated unlearning networks (FUN). The main contributions of this paper are summarized as follows:
\begin{itemize}
    \item We establish a system model for FUN that jointly captures the convergence dynamics of the distributed approximate Newton-type calibration process, local computation at resource-constrained edge devices, and a worst-case robust transmission design under bounded CSI errors. We derive a closed-form lower bound on the global calibration rounds that explicitly characterizes the fundamental trade-off between local optimization accuracy and communication overhead.
    
    \item We develop an efficient iterative algorithm to solve the formulated non-convex delay minimization problem. By exploiting the proved monotonicity and convexity properties of the feasibility region, we decompose the problem via a uniform scan over the local accuracy parameter, within which the optimal delay, bandwidth, transmit power, and central processing unit (CPU) frequency are determined through a nested bisection and golden-section search framework. We rigorously prove that the proposed algorithm achieves polynomial-time complexity.
    
    \item We demonstrate through extensive numerical simulations that the proposed algorithm significantly reduces the unlearning completion time compared to baseline schemes. The results further reveal that joint optimization of communication and computation resources provides substantial robustness against channel uncertainty and stringent energy constraints, validating the practical efficacy of the proposed framework in wireless edge environments.
\end{itemize}

The remainder of this paper is organized as follows. Section \ref{sec.sm} describes the system model. Section \ref{sec.radm} formulates the delay minimization problem and presents the proposed algorithm. Section \ref{sec.sr} provides numerical results, and Section \ref{sec.c} concludes the paper.

The main notations used in the paper are summarized in Table \ref{tab:notations}.
\begin{table}[!t]
\caption{List of Main Notations}
\label{tab:notations}
\centering
\footnotesize
\setlength{\tabcolsep}{1.5pt}
\renewcommand{\arraystretch}{1.15}
\begin{tabular}{l p{6.3cm}}
\toprule
\textbf{Notation} & \textbf{Description} \\
\midrule
$K$ & Number of users \\
$k_u$ & Index of the unlearning user \\
$D_k$ & Number of local data samples \\
$I$ & Number of global FL rounds \\
$J(\eta)$ & Number of global FU rounds \\
$\epsilon_0$ & Target global accuracy \\
$\eta$ & Local optimization accuracy \\
\midrule
$\boldsymbol{w}_i$ & Global model at the $i$-th FL round \\
$\boldsymbol{\widetilde{w}}^{(j)}$ & Global model at the $j$-th FU round \\
$D^{\text{cali}}$ & Total calibration data samples \\
$\boldsymbol{h}_k^{(j)}$ & Historical update of user $k$ at round $j$ \\
$\boldsymbol{\hat{h}}_k^{(j)}$ & Calibration update of user $k$ at round $j$ \\
$\boldsymbol{\widetilde{h}}_k^{(j)}$ & Calibrated update at the BS \\
$\alpha,\beta$ & Lower and upper calibration scaling bounds \\
$\gamma,L$ & Strong convexity and Lipschitz constants \\
\midrule
$g_k$ & Real channel gain of user $k$ \\
$\hat{g}_k$ & Estimated channel gain at the BS \\
$\epsilon_k$ & Channel estimation error bound \\
$B$ & Total system bandwidth \\
$b_k$ & Bandwidth allocated to user $k$ \\
$p_k$ & Transmit power of user $k$ \\
$N_0$ & Noise power spectral density \\
$r_k$ & Achievable uplink transmission rate \\
$s$ & Upload data size (bits) \\
$t_k$ & Transmission time per round \\
\midrule
$f_k$ & CPU frequency of user $k$ \\
$C_k$ & CPU cycles per sample \\
$\kappa$ & Effective switched capacitance \\
$A_k$ & Computation coefficient, $A_k=vC_kD_k$ \\
$\tau_k$ & Local computation time per round \\
$E_k$ & Total energy consumption of user $k$ \\
\midrule
$T$ & Total unlearning delay \\
$T_k$ & Per-user completion delay \\
\bottomrule
\end{tabular}
\end{table}

\section{System Model}\label{sec.sm}
\begin{figure}[t]
    \centering
    \includegraphics[width=1\linewidth]{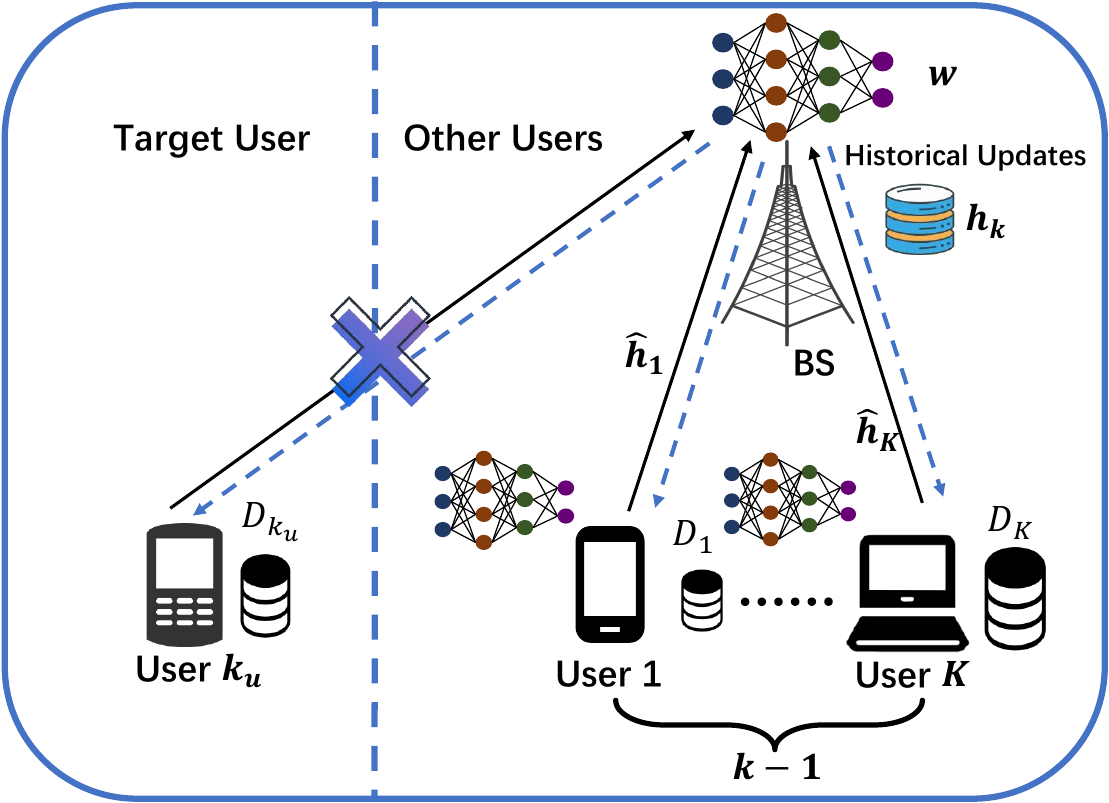}
    \caption{FU over wireless networks.}
    \label{model}
\end{figure}

Consider a wireless edge network consisting of a base station (BS) and a set of $K$ users, denoted by $\mathcal{K} = \{1, 2, \dots, K\}$. Each user $k \in \mathcal{K}$ has a local dataset $\mathcal{D}_k=\{x_{kl},y_{kl}\}_{l=1}^{D_k}$ with $D_k$ data samples, where $x_{kl}\in \mathbb{R}^d$ is an input vector and $y_{kl}$ is its corresponding output. In the considered FUN, a specific user $k_u \in \mathcal{K}$ requests to remove its data contribution from the globally trained model, as shown in Fig.~\ref{model}. To achieve this, the remaining clients, denoted by the set $\tilde{\mathcal{K}} = \mathcal{K} \setminus \{k_u\}$, collaboratively execute a calibration process coordinated by the BS. 

\subsection{FL Model}
\subsubsection{Architecture of FL}
Following the distributed approximate Newton (DANE) framework proposed in \cite{56},  the users collaboratively train a unified global model under the coordination of the BS. Let $I$ denote the total number of global rounds required to achieve a target global accuracy $\epsilon_0$. We assume that $\boldsymbol{w}_i$ represents the global model at the $i$-th FL round. At the $i$-th round ($i = 1, 2, \ldots, I$), each user $k$ trains the received model on its local dataset $\mathcal{D}_k$ for $I_{\text{local}}$ epochs according to local accuracy $\eta$, producing an update $\boldsymbol{h}_k^i$ with respect to $\boldsymbol{w}_i$. The BS aggregates the received updates $\{\boldsymbol{h}_1^i, \boldsymbol{h}_2^i, \ldots, \boldsymbol{h}_K^i\}$ to obtain the new global model $\boldsymbol{w}_{i+1}$, which will be used in the next round.

When the global accuracy $\epsilon_0$ is satisfied, the BS stops the above iterative training process and get the final FL model $w^{I}$.

\subsubsection{Storage of Historical Updates}
To meet the requirements of FU, the BS stores the historical updates of all users during the FL phase. Let $\Delta i$ denote the storage interval (i.e., the server records an update every $\Delta i$ global rounds). The stored updates for user $k$ are denoted as $\boldsymbol{h}_k^{(j)}$ for $j = 1,2,\dots,J$, where $\boldsymbol{h}_k^{(j)}=\boldsymbol{h}_k^{1+(j-1)\Delta i}$, and $J=\lfloor\frac{I}{\Delta i}\rfloor$ denotes the total number of stored rounds and the theoretical lower bound of global iterations in the unlearning phase.

We assume that the global accuracy of FU is also $\epsilon_0$ and the local accuracy is $\eta$. Consequently, $J$ is mathematically governed by both $\epsilon_0$ and $\eta$. The relationship among $J$, $\epsilon_0$, and $\eta$ will be discussed in the FU Model of Section~\ref{sec.sm}.

\subsection{FU Model}

In the FU phase, we incorporate FedEraser proposed in \cite{33} within the DANE-based FL framework, which is summarized in Algorithm~\ref{alg1}. We assume that $\boldsymbol{\widetilde{w}}^{(j)}$ is the global model of $j$-th round of FU. Since the dataset of user $k$ is $\mathcal{D}_k$, the total loss function of user $k$ except $k_u$ is
\begin{equation}\label{eq1}
F_k(\boldsymbol{\widetilde{w}}^{(j)})=\frac{1}{D_k}\sum_{l=1}^{D_k}f(\boldsymbol{\widetilde{w}}^{(j)},x_{kl},y_{kl}), \quad k \in \mathcal{\widetilde{K}}.
\end{equation}
The FU calibration problem can be formulate as follows
\begin{align}\label{eq2}
\min_{\boldsymbol{\widetilde{w}}^{(j)}} F(\boldsymbol{\widetilde{w}}^{(j)})&=\sum_{k \in \mathcal{\widetilde{K}}}\frac{D_k}{D^{cali}}F_k(\boldsymbol{\widetilde{w}}^{(j)})\nonumber\\
&=\frac{1}{D^{cali}}\sum_{k \in \mathcal{\widetilde{K}}}\sum_{l=1}^{D_k}f(\boldsymbol{\widetilde{w}}^{(j)},\boldsymbol{x}_{kl},y_{kl}),
\end{align}
where $D^{cali}=\sum_{k \in \mathcal{\widetilde{K}}}D_k$ is the total data samples of all users $k \in \mathcal{\widetilde{K}}$.

For the local update phase, each user $k$ except $k_u$ computes the local optimization problem
\begin{align}\label{eq3}
\min_{\boldsymbol{\hat{h}}_k \in \mathbb{R}^d} \quad  G_k&\left(\boldsymbol{\widetilde{w}}^{(j)},\boldsymbol{\hat{h}}_k\right)\triangleq F_k\left(\boldsymbol{\widetilde{w}}^{(j)}+\boldsymbol{\hat{h}}_k\right)\nonumber\\&-\left(\nabla F_k\left(\boldsymbol{\widetilde{w}}^{(j)}\right)-\xi\nabla F\left(\boldsymbol{\widetilde{w}}^{(j)}\right)\right)^T \boldsymbol{\hat{h}}_k.
\end{align}
In problem (3), $\xi$ is a constant value and $\boldsymbol{\hat{h}}_k$ denotes the calibration update for FU. To solve the local optimization problem (3), we use the gradient method
\begin{align}\label{eq4}
\boldsymbol{\hat{h}}_k^{(j),(n+1)}=\boldsymbol{\hat{h}}_k^{(j),(n)}-\delta   \nabla G_k\left(\boldsymbol{\widetilde{w}}^{(j)},\boldsymbol{\hat{h}}_k^{(j),(n)}\right),
\end{align}
where $\delta$ is the step size, and $\boldsymbol{\hat{h}}_k^{(j),(n)}$ is the value of $\boldsymbol{\hat{h}}_k$ at $n$-th local iteraion with given vector $\boldsymbol{\widetilde{w}}^{(j)}$. 
To ensure a fair comparison with the FL retraining baseline, we also use $\eta$ as the accuracy of FU local problems. According to \cite{56}, we have
\begin{align}\label{eq5}
G_k&(\boldsymbol{\widetilde{w}}^{(j)},\boldsymbol{\hat{h}}_k^{(j),(n)})-G_k(\boldsymbol{\widetilde{w}}^{(j)},\boldsymbol{\hat{h}}_k^{(j)*})\nonumber\\
&\leq \eta\left(G_k(\boldsymbol{\widetilde{w}}^{(j)},0)-G_k(\boldsymbol{\widetilde{w}}^{(j)},\boldsymbol{\hat{h}}_k^{(j)})\right),
\end{align}
where $\boldsymbol{\hat{h}}_k^{(j)*}$ is the actual optimal solution of problem (3).
In Algorithm~\ref{alg1}, the solution $\boldsymbol{w}^{(j)}$ with the accuracy $\epsilon_0$ means that
\begin{equation}\label{eq8}
F(\boldsymbol{\widetilde{w}}^{(j)})-F(\boldsymbol{\widetilde{w}}^*)\leq\epsilon_0\left(F(\boldsymbol{\widetilde{w}}^{(0)})-F(\boldsymbol{\widetilde{w}}^*)\right),
\end{equation}
where $\boldsymbol{\widetilde{w}}^*$ is the actual optimal solution.

\begin{algorithm}[t]
    \caption{FUN Algorithm}\label{alg1}
    \begin{algorithmic}[1]
    \STATE Initialize global model $\boldsymbol{ \widetilde{w}}^0 = \boldsymbol{w}^0$ and global iteration number $j = 0$.
     \REPEAT
            \STATE Each user $k$ except $k_u$ computes $\nabla F_k(\boldsymbol{\widetilde{w}}^{(j)})$ and sends it to the BS.
            \STATE The BS computes $\nabla F(\boldsymbol{\widetilde{w}}^{(j)}) = \frac{1}{K-1}\sum_{k \in \mathcal{\widetilde{K}}} \nabla F_k(\boldsymbol{\widetilde{w}}^{(j)})$, which is broadcast to all users except user $k_u$.
            \FOR{user $k \in \mathcal{\widetilde{K}}$  \textbf{in parallel}}
            \STATE Initialize the local iteration number $n = 0$ and set $\boldsymbol{\hat{h}}_k^{(j),(0)}=\boldsymbol{0}$.
            \REPEAT
                \STATE Update $\boldsymbol{\hat{h}}_k^{(j),(n+1)} = \boldsymbol{\hat{h}}_k^{(j),(n)}-\delta \nabla G_k(\boldsymbol{\widetilde{w}}^{(j)},\boldsymbol{\hat{h}}_k^{(j),(n)})$ and $n=n+1$.
            \UNTIL{the accuracy $\eta$ of locall optimization problem is obtained.}
            \STATE Denote $\boldsymbol{\hat{h}}_k^{(j)}=\boldsymbol{\hat{h}}_k^{(j),(n)}$ and each user except $k_u$ sends $\boldsymbol{\hat{h}}_k^{(j)}$ to the BS.  
        \ENDFOR
    \STATE The BS computes the calibration of $\boldsymbol{h}_k^{(j)}$ as
    \begin{align}\label{eq6}
        \boldsymbol{\widetilde{h}}_k^{(j)} = \left\| \boldsymbol{h}_k^{(j)}\right\|\frac{\boldsymbol{\hat{h}}_k^{(j)}}{\left\|\boldsymbol{\hat{h}}_k^{(j)}\right\|}.
    \end{align}
    \STATE Then the BS can computes  
    \begin{align}\label{eq7}
    \boldsymbol{\widetilde{w}}^{(j+1)} = \boldsymbol{\widetilde{w}}^{(j)}+\frac{1}{K-1}\sum_{k \in \mathcal{\widetilde{K}}} \boldsymbol{\widetilde{h}}_k^{(j)},
    \end{align}
    and broadcasts the value to all users except user $k_u$.
    \STATE Set $j=j+1$.
\UNTIL{the accuracy $\epsilon_0$ of problem \eqref{eq2} is obtained}.
\end{algorithmic}
\end{algorithm}

In Algorithm~\ref{alg1}, $ \boldsymbol{\widetilde{h}}_k^{(j)}$ represents the update after BS calibration. For convenience in the subsequent derivation, we assume that there exist positive real numbers $\alpha=\min_{k,j}\frac{\|\boldsymbol{h}_k^{(j)}\|}{\|\boldsymbol{\hat{h}}_k^{(j)}\|}$ and $\quad
\beta=\max_{k,j}\frac{\|\boldsymbol{h}_k^{(j)}\|}{\|\boldsymbol{\hat{h}}_k^{(j)}\|}$, where the minimum and maximum are taken over all $k \in \mathcal{\widetilde{K}}$ and $j = 1,2,\dots,J$. Then, the following inequality holds
\begin{align}\label{eq9}
\alpha\left\|\boldsymbol{\hat{h}}_k^{(j)}\right\|\leq\left\|\boldsymbol{h}_k^{(j)}\right\|\leq \beta\left\|\boldsymbol{\hat{h}}_k^{(j)}\right\|,
\end{align}
where $\alpha$ and $\beta$ is determined by the model parameters.
In order to analyze the convergence of Algorithm~\ref{alg1}, we assume that $F_k(\boldsymbol{w})$ is $L$-Lipschitz continuous and $\gamma$-strongly convex,i.e.,
\begin{equation}\label{eq10}
\gamma \boldsymbol{I}\preceq \nabla^2F_k(\boldsymbol{w})\preceq L\boldsymbol{I}, \forall k \in \mathcal{\widetilde{K}},
\end{equation}
where $\gamma$ and $L$ are assumed to be known constants. Under \eqref{eq9} and \eqref{eq10}, we can get the relationship between the lower bound of global iterations $J$ and the local accuracy $\eta$.

\textit{Lemma 1: }if we run Algorithm 1 with $0<\xi\leq \frac{\gamma}{L\beta}$ for
\begin{equation}\label{eq11}
j \geq \frac{a}{(\gamma-L\xi\beta)(\sqrt{\gamma}-\sqrt{\eta L})^2+(1-\eta)\gamma^2}\triangleq J(\eta)
\end{equation}
iterations with $a=\frac{2L^2\mathrm{ln}(1/\epsilon_0)}{\xi\beta}$, we have $F(\boldsymbol{\tilde{w}}^{(j)})-F(\boldsymbol{\tilde{w}}^*)\leq \epsilon_0(F(\boldsymbol{\tilde{w}}^{(0)})-F(\boldsymbol{\tilde{w}}^*))$.

\textit{Proof: }See Appendix A.\hfill $\blacksquare$

To get the iterations for local problem with a local accuracy $\eta$, we set $v = \frac{2}{(2-L\delta)\delta\gamma}$. Since the update calibration is performed at the BS, the number of local iterations can refer to the derivation in \cite{56}. 

\textit{Lemma 2: }If we set $\delta < \frac{2}{L}$ and run the gradient method for $n\geq v\mathrm{log}_2(1/\eta)$ iterations at each user $k \in \mathcal{\widetilde{K}}$, we can solve local problem with an accuracy $\eta$.

\textit{Proof: }See Appendix A of \cite{56}.\hfill $\blacksquare$

In the following,  we respectively use $J(\eta)$ and $v\mathrm{log}_2(1/\eta)$ to approximate the number of global and local iterations.

\subsection{Computation and Transmission Model}
\subsubsection{Local Computation}
Denote the computation capacity of user $k$ by $f_k$ (cycles/second), which is measured by the number of CPU cycles per second. Then, the computation time of user $k$ is
\begin{equation}\label{eq12}
\tau_k=\frac{A_k\mathrm{log}_2(1/\eta)}{f_k}, \forall k \in \mathcal{\widetilde{K}}
\end{equation}
where $A_k = v C_k D_k$ and $C_k$ (cycles/sample) is the number of CPU cycles necessary for computing one sample data at user $k$. According to Lemma 1 in \cite{57}, the computation energy per round is given by 
\begin{equation}\label{eq13}
E_k^{\text{comp}}  = \kappa f_k^2 A_k \log_2\left(\frac{1}{\eta}\right),
\end{equation}
where $\kappa$ is the effective switched capacitance coefficient of user $k$'s processor. 

\subsubsection{Wireless Transmission}

After local computation, all users upload their update to the BS via frequency domain multiple access (FDMA). 
In practical wireless communication systems, the CSI acquired at the BS is often imperfect due to channel estimation errors and feedback delays. To explicitly capture the CSI uncertainty, we adopt a bounded uncertainty model as in \cite{58}.

Specifically, the relation between the obtained CSI and the real CSI is usually modeled as 
\begin{equation}\label{eq14}
g_k = \hat{g}_k + \Delta g_k,\quad \forall k \in \mathcal{\widetilde{K}},
\end{equation}
where $\hat{g}_k$ denotes the channel estimate available at the BS, and $\Delta g_k$ represents the channel estimation error. We assume that the error is bounded within a known uncertainty region $\left|\Delta g_k\right|^2\leq \epsilon_k^2$, where $\epsilon_k>0$ determines the size of the uncertainty region.  

Under this model, the achievable uplink transmission rate of user $k$ is given by 
\begin{equation}\label{eq15}
r_k = b_k \log_2 \left( 1 + \frac{\left|g_k\right|^2 p_k}{N_0 b_k} \right),\quad\forall k \in \mathcal{\widetilde{K}},
\end{equation}
where $b_k$ is the bandwidth allocated to user $k$, $p_k$ is the transmit power of user $k$, and $N_0$ is the power spectral density of the Gaussian noise. Due to the limited bandwidth, we have $\sum_{k\in \mathcal{K}}^K b_k \leq B$, where $B$ is the total bandwidth.
To guarantee reliable communication under channel uncertainty, we adopt a worst-case robust design. Specifically, we ensure that the transmission requirement is satisfied for all possible channel realizations within the uncertainty region.
Accordingly, the following robust constraint must hold
\begin{equation}\label{eq16}
    t_k b_k \log_2 \left( 1 + \frac{(\left| \hat{g}_k\right|-\epsilon_k)^2p_k}{N_0 b_k} \right) \geq s, \quad \forall k \in \mathcal{\widetilde{K}},
\end{equation}
where $s$ is the data size, and $(\left| \hat{g}_k\right|-\epsilon_k)^2$ corresponds to the worst-case channel condition.

The energy consumed for uploading the local update within a time duration $t_k$ is 
\begin{equation}\label{eq17}
    E_k^{\text{comm}} = p_k t_k.
\end{equation} 

\subsubsection{Information Broadcast}
In this step, the BS aggregates the updates $\boldsymbol{\hat{h}}_k$ from users and then broadcasts the global model $\boldsymbol{\widetilde{w}}^{(j)}$ to all users in the downlink. Due to the high power of the BS and large downlink bandwidth, we ignore the downlink time.

The delay of each user $k\in \widetilde{\mathcal{K}}$ includes the local computation time and transmit time. Base on (\ref{eq12}) and (\ref{eq16}), the delay $T_k$ of user k will be
\begin{equation}\label{eq18}
T_k=J(\eta)\left(\tau_k+t_k\right)=J(\eta)\left(\frac{A_k\mathrm{log}_2(1/\eta)}{f_k}+t_k\right).
\end{equation}
We define $T=\max_{k\in{\mathcal{\widetilde{K}}}}T_k$ as the delay for FU algorithm and we have $T_k\leq T$. Base on (\ref{eq13}) and (\ref{eq17}), the total energy consumption of user $k$ will be
\begin{align}\label{eq19}
E_k &= J(\eta)(E_k^\text{comp}+E_k^\text{comm}) \nonumber\\
&=J(\eta)\left( \kappa f_k^2 A_k \log_2\left(\frac{1}{\eta}\right) + p_k t_k \right), \quad \forall k \in \widetilde{\mathcal{K}}.
\end{align}
Due to the limited battery resources of user edge devices, there exists an energy consumption upper bound $E_k^{max}$.
We have $E_k\leq E_k^{max}$.

\section{Resource Allocation for Delay minimization}\label{sec.radm}
In this section, we formulate the delay minimization problem for FUN. To solve the non-convex optimization problem, we propose an iterative algorithm. 

\subsection{Problem Formulation}
We now pose the joint resource allocation problem aimed at minimizing the total time of the FU process
\begin{subequations}\label{eq20}
\begin{align}
\min_{T, \boldsymbol{t}, \boldsymbol{b},\boldsymbol{f},\boldsymbol{p},\eta} \quad & T \tag{\ref{eq20}}\\
\textrm{s.t.} \quad\quad& J(\eta)\left(\frac{A_k\log_2(1/\eta)}{f_k}+t_k\right)\leq T, \quad \forall k \in \widetilde{\mathcal{K}}, \label{eq20a}\\
&J(\eta)\left( \kappa f_k^2 A_k \log_2(\frac{1}{\eta}) + p_k t_k \right) \leq E_k^{\max}, \forall k \in \widetilde{\mathcal{K}}, \label{eq20b}\\
&t_k b_k \log_2 \left( 1 + \frac{(\left| \hat{g}_k\right|-\epsilon_k)^2p_k}{N_0 b_k} \right) \geq s, \forall k \in \widetilde{\mathcal{K}}, \label{eq20c}\\
&\sum_{k \in \widetilde{\mathcal{K}}} b_k\leq B, \quad \forall k \in \widetilde{\mathcal{K}}, \label{eq20d}\\
&0\leq f_k \leq f_k^{\max},\quad 0\leq p_k\leq p_k^{\max}, \quad \forall k \in \widetilde{\mathcal{K}}, \label{eq20e}\\
&0\leq \eta \leq 1, \label{eq20f}\\
&t_k\geq 0,\; b_k\geq 0, \quad \forall k \in \widetilde{\mathcal{K}}, \label{eq20g}
\end{align}
\end{subequations}
where $\boldsymbol{t} = [t_1,\dots, t_{k_u-1}, t_{k_u+1},\dots, t_K]^T$, $\boldsymbol{b} = [b_1,\dots, b_{k_u-1}, b_{k_u+1},\dots, b_K]^T$, $\boldsymbol{f} = [f_1,\dots, f_{k_u-1}, f_{k_u+1},\dots, f_K]^T$, and $\boldsymbol{p} = [p_1,\dots, p_{k_u-1}, p_{k_u+1},\dots, p_K]^T$.
$f_k^{\max}$ and $p_k^{\max}$ are the maximum computation capacity and maximum transmit power of user $k$, respectively.  
Constraint \eqref{eq20a} ensures that the total time for each participating user does not exceed the overall unlearning delay $T$.  
Constraint \eqref{eq20b} limits the total energy consumption per user, including both computation and transmission, to be within the user's maximum energy budget $E_k^{\max}$. 
Constraint \eqref{eq20c} ensure that users are able to upload data of size $s$
within transmit time $t_k$ under the worst channel conditions.  
Constraint \eqref{eq20d} is the total bandwidth limit.  
Constraints \eqref{eq20e} specify the feasible ranges for the CPU frequency and transmit power of each user.  
Constraint \eqref{eq20f} defines the allowed range of the local optimization accuracy $\eta$.  
Finally, constraints \eqref{eq20g} enforce non‑negativity of the transmission time and bandwidth allocation.

Problem \eqref{eq20} is a non-convex problem and is hard to solve directly. The main challenge comes from the coupling of variables, such as time $t_k$ , bandwidth $b_k$ , and power $p_k$ , in the rate constraints. Also, the max operator in the objective function and the variable $\eta$ make the problem complex. To address these challenges, we decompose the problem into a series of subproblems.

\subsection{Feasibility Analysis for a Single User}
For the fixed $\eta$, $T$, and bandwidth $b_k$ allocated to user $k$, the feasibility of user $k$ is determined by the existence of transmission time $T$ and CPU frequency $f_k$ that satisfy constraints \eqref{eq20a}, \eqref{eq20b}, \eqref{eq20c}, and \eqref{eq20e}. For notational convenience, we define
\begin{equation}\label{eq21}
Q(\eta) = J(\eta) \log_2\left(\frac{1}{\eta}\right).
\end{equation}
According to the delay constraint (\ref{eq20a}), we obtain
\begin{equation}\label{eq22}
f_k \ge f_k^{\min}(t_k) \triangleq \frac{Q(\eta) A_k}{T - J(\eta) t_k},
\end{equation}
where $t_k \in (0, T/J(\eta))$. From the transmission constraint (\ref{eq20c}), we obtain
\begin{equation}\label{eq23}
p_k \ge p_k^{\min}(t_k) \triangleq \frac{N_0 b_k}{(\left| \hat{g}_k\right|-\epsilon_k)^2}\left(2^{\frac{s}{t_k b_k}} - 1\right).
\end{equation}
By substitute $f_k^{\min}(t_k)$ and $p_k^{\min}(t_k)$ into the energy constraint (\ref{eq20b}), we can obtain
\begin{equation}\label{eq24}
\kappa \bigl(f_k^{\min}(t_k)\bigr)^2 Q(\eta) A_k + J(\eta)\, p_k^{\min}(t_k)\, t_k \le E_k^{\max}.
\end{equation}
Denote the left-hand side of \eqref{eq24} as $E_k^{\text{total}}(t_k)$.

For given $\eta$, $T$, and $b_k$, the user $k$ is feasible if and only if there exists $t_k \in (0, T/J(\eta))$ such that
\begin{equation}\label{eq25}
f_k^{\min}(t_k) \le f_k^{\max},\quad p_k^{\min}(t_k) \le p_k^{\max},\quad E_k^{\text{total}}(t_k) \le E_k^{\max}.
\end{equation}
Then, base on $f_k^{\min}(t_k) \le f_k^{\max}$, we can obtain the upper bound of $t_k$, denoted as
\begin{equation}\label{eq26}
    t_k^{\max} = \frac{T}{J(\eta)}-\frac{A_k\log_2(1/\eta)}{f_k^{max}}.
\end{equation}
Base on $p_k^{\min}(t_k) \le p_k^{\max}$, we can obtain the lower bound of $t_k$, denoted as
\begin{equation}\label{eq27}
    t_k^{\min} = \frac{s}{b_k \log_2 \left( 1 + \frac{(\left| \hat{g}_k\right|-\epsilon_k)^2p_k^{max}}{N_0 b_k} \right)}.
\end{equation}

\textit{Lemma 3: }Under the condition of $t_k^{\min} \le t_k^{\max}$, the feasibility of user $k$ is determined by evaluating the optimal transmission time $t_k^*\in [t_k^{\min},t_k^{\max}]$ found via the golden-section search. If the total energy $E_{\text{total}}(t_k^*)$ is less than or equal to the energy budget $E_k^{\max}$, problem \eqref{eq25} is feasible.

\textit{Proof: }To find feasible $t_k$, we analyze the properties of $f_k^{\min}(t_k)$, $p_k^{\min}(t_k)$, and $E_k^{\text{total}}(t_k)$. We know that $f_k^{\min}(t_k)$ is a convex and increasing function and $p_k^{\min}(t_k)$ is a convex and decreasing function. It is easy to proof that $(f_k^{\min}(t_k))^2 = \left(\frac{Q(\eta) A_k}{T - J(\eta)t_k}\right)^2$ is strictly convex and increasing with respect to $t_k$, and $p_k^{\min}(t_k) t_k = \frac{N_0 b_k t_k}{(\left| \hat{g}_k\right|-\epsilon_k)^2} (2^{\frac{s}{t_k b_k}} - 1)$ is a convex and decreasing function of $t_k$ for $t_k > 0$. The total energy function $E_k^{\text{total}}(t_k)$ is the sum of two convex functions, so $E_k^{\text{total}}(t_k)$ is a convex function. This property guarantees that the optimal $t_k^*$ minimizing energy consumption can be efficiently found using a Golden-section search.

The procedure is given in Algorithm \ref{alg:feasibility_single}. 
\begin{algorithm}[ht]
\caption{Feasibility Test for Single User $k$}
\label{alg:feasibility_single}
\begin{algorithmic}[1]
\REQUIRE $\eta$, $T$, $b_k$, $E_k^{\max}$, Tolerance $\epsilon_t > 0$.
\ENSURE Feasible (True/False), Optimal $t_k^*$.
\STATE Compute $J(\eta)$ and $Q(\eta) = J(\eta) \log_2(1/\eta)$.
\STATE Calculate time bounds $t_k^{\min}$ via \eqref{eq27} and $t_k^{\max}$ via \eqref{eq26}.
\IF{$t_k^{\min} \ge t_k^{\max}$}
    \RETURN \textbf{False} 
\ENDIF
\STATE Set $t_a = t_k^{\min}$, $t_b = t_k^{\max}$.
\REPEAT
    \STATE $t_1 = t_a + 0.382(b-a)$, $t_2 = t_a + 0.618(b-a)$.
    \STATE Compute $f_k^{\min}(t_1)$ and $p_k^{\min}(t_1)$.
    \STATE Compute Total Energy $E_{\text{total}}(t_1)$ via \eqref{eq24}.
    \STATE Compute $f_k^{\min}(t_2)$ and $p_k^{\min}(t_2)$.
    \STATE Compute Total Energy $E_{\text{total}}(t_2)$ via \eqref{eq24}.
    \IF{$E_{\text{total}}(t_1) < E_{\text{total}}(t_2)$}
        \STATE $t_b = t_2$.
    \ELSE
        \STATE $t_a = t_1$.
    \ENDIF
\UNTIL{$|t_b - t_a| \le \epsilon_t$}
\STATE $t_k^* = (t_a + t_b)/2$.
\STATE Compute final energy $E_{\text{total}}(t_k^*)$.
\IF{$E_{\text{total}}(t_k^*) \le E_k^{\max}$}
    \RETURN \textbf{True}, $t_k^*$.
\ELSE
    \RETURN \textbf{False}.
\ENDIF
\end{algorithmic}
\end{algorithm}
\subsection{Minimum Required Bandwidth}

For a fixed local accuracy $\eta$ and target delay $T$, the feasibility of user $k$ exhibits a strict monotonic structure with respect to the allocated bandwidth $b_k$. This structural property is pivotal as it transforms the non-convex feasibility problem into an efficient root-finding process. Specifically, we prove that if a user is feasible at a given bandwidth $b_k$, it remains feasible for any larger bandwidth $b_k' \ge b_k$. A larger bandwidth inherently increases the channel capacity, thereby relaxing the transmission delay constraint and reducing the required transmit power for a fixed data size.

Formally, the monotonicity is established in the following lemma.

\textit{Lemma 4: }
For fixed $\eta$ and $T$, if user $k$ is feasible at bandwidth $b_k$, then it is also feasible at any bandwidth $b_k' \ge b_k$.

\textit{Proof: }
See Appendix B. \hfill $\blacksquare$

Based on Lemma 4, the minimum bandwidth $b_k^{\min}$ required for user $k$ to be feasible is the smallest $b_k \in [0, B]$ such that the conditions in \eqref{eq25} are satisfied. This value can be efficiently computed via a bisection search over the interval $[0, B]$. The convergence rate of this bisection method is linear with respect to the search tolerance $\epsilon_b$, requiring $\mathcal{O}(\log(1/\epsilon_b))$ iterations to achieve a solution precision of $\epsilon_b$. 

The sensitivity of $b_k^{\min}$ to the channel uncertainty bound $\epsilon_k$ is critical. As derived in the robust rate constraint \eqref{eq16}, a larger uncertainty bound forces the system to operate under a more conservative channel gain. Consequently, to maintain feasibility under stricter robustness constraints, the algorithm must allocate a larger $b_k^{\min}$ or compensate with higher transmit power, highlighting the trade-off between robustness and spectral efficiency. The procedure for computing $b_k^{\min}$ is summarized in Algorithm~\ref{alg:min_bw}.
\begin{algorithm}[ht]
\caption{Compute $b_k^{\min}$ for user $k$}
\label{alg:min_bw}
\begin{algorithmic}[1]
\REQUIRE $\eta$, $T$, tolerance $\epsilon_b>0$.
\STATE Set $b_{\text{low}} = 0$, $b_{\text{high}} = B$.
\REPEAT
    \STATE $b = (b_{\text{low}} + b_{\text{high}})/2$.
    \STATE Verify the feasibility of problem \eqref{eq25}  with $b_k = b$.
    \IF{feasible}
        \STATE $b_{\text{high}} = b$.
    \ELSE
        \STATE $b_{\text{low}} = b$.
    \ENDIF
\UNTIL{$b_{\text{high}} - b_{\text{low}} \le \epsilon_b$} 
\RETURN $b_k^{\text{min}}=b_{\text{high}}$.
\end{algorithmic}
\end{algorithm}

\subsection{Delay Minimization for a Given Accuracy}
For a fixed local accuracy $\eta$, the global feasibility condition is governed by the bandwidth constraint. Specifically, the sum of the minimum required bandwidths for all remaining clients must not exceed the total available bandwidth $B$, which is mathematically expressed as
\begin{equation}\label{eq29}
\sum_{k\in\tilde{\mathcal{K}}} b_k^{\min}(\eta, T) \le B.
\end{equation}

To efficiently solve this feasibility problem, we exploit the inherent monotonicity property with respect to the delay $T$. The following lemma establishes this critical structural property.

\textit{Lemma 5: }
The feasibility of problem~\eqref{eq20} exhibits a strict monotonic structure with respect to $T$. If problem~\eqref{eq20} is feasible for a given $T$, then it remains feasible for any $T' > T$. Conversely, if it is infeasible for $T$, then it is also infeasible for any $T' < T$.

\textit{Proof: } See Appendix E in \cite{56}. \hfill $\blacksquare$

Lemma 5 allows us to employ a bisection search to find the minimum feasible delay $T^*(\eta)$. The convergence rate of this method is logarithmic with respect to the inverse of the tolerance $\epsilon_T$, specifically $\mathcal{O}(\log(1/\epsilon_T))$. 

We initialize the search interval as $[T_{\min}, T_{\max}]$. Here, $T_{\min}=0$ serves as the trivial lower bound. The upper bound $T_{\max}$ is chosen to be sufficiently large to guarantee feasibility. The procedure for finding $T^*(\eta)$ is outlined in Algorithm~\ref{alg:delay_bisection}.

\begin{algorithm}[ht]
\caption{Find minimal delay $T^*(\eta)$ for fixed $\eta$}
\label{alg:delay_bisection}
\begin{algorithmic}[1]
\REQUIRE $\eta$, total bandwidth $B$, tolerance $\epsilon_T>0$.
\STATE Compute $J(\eta)$ from \eqref{eq10} and $Q(\eta)$ from \eqref{eq21}.
\STATE Set $T_{\text{low}} = 0$, $T_{\text{high}} = T_{\max}$.
\REPEAT
    \STATE $T = (T_{\text{low}} + T_{\text{high}})/2$.
    \STATE For each user $k\in\tilde{\mathcal{K}}$, compute $b_k^{\min}$ via Algorithm~\ref{alg:min_bw}.
    \IF{$\sum_{k} b_k^{\min} \le B$}
        \STATE $T_{\text{high}} = T$.
    \ELSE
        \STATE $T_{\text{low}} = T$.
    \ENDIF
\UNTIL{$T_{\text{high}} - T_{\text{low}} \le \epsilon_T$} 
\RETURN $T^*(\eta) = T_{\text{high}}$.
\end{algorithmic}
\end{algorithm}

\subsection{Optimization over Local Accuracy}

The selection of the local accuracy $\eta$ plays a pivotal role in determining the overall completion time $T^*(\eta)$, exhibiting a non-trivial trade-off between local computation and global communication.

Specifically, the local accuracy $\eta$ influences the number of global iterations $J(\eta)$ and the per-round computational load $Q(\eta) = J(\eta) \log_2(1/\eta)$. On one hand, a larger $\eta$ reduces the computational effort $Q(\eta)$ required per round, thereby decreasing the local computation time $\tau_k$. However, it inevitably increases the number of global calibration rounds $J(\eta)$ required for the FUN algorithm to converge, leading to higher cumulative communication overhead.
On the other hand, a smaller $\eta$ reduces $J(\eta)$ by shifting the workload towards local computation, but this comes at the cost of significantly increasing the per-round transmission time $t_k$ due to the larger volume of data that needs to be uploaded.

To address this, we resort to a one-dimensional uniform sampling strategy over the feasible domain of $\eta$. We discretize the interval $[10^{-3}, 0.99]$ into $M$ equally spaced points $\{\eta^{(1)}, \eta^{(2)}, \dots, \eta^{(M)}\}$. For each candidate $\eta^{(m)}$, we invoke Algorithm~\ref{alg:delay_bisection} to compute the corresponding minimal system delay $T^*(\eta^{(m)})$. The optimal local accuracy $\eta^*$ is then selected as the one that minimizes the resulting delay
\begin{equation}
    \eta^* = \arg \min_{\eta^{(m)}} T^*(\eta^{(m)}), \quad m = 1, \dots, M.
\end{equation}
The overall optimization is summarized in Algorithm~\ref{alg:overall_simple}.

\begin{algorithm}[ht]
\caption{Delay Minimization for FUN}
\label{alg:overall_simple}
\begin{algorithmic}[1]
\REQUIRE System parameters $\{A_k, f_k^{\max}, p_k^{\max}, E_k^{\max}, \hat{g}_k, \epsilon_k\}$, $s$, $B$, $N_0$, number of sampling points $M$.
\STATE Define $\eta_{\min}=10^{-3}$, $\eta_{\max}=0.99$, and step $\Delta\eta = (\eta_{\max}-\eta_{\min})/(M-1)$.
\STATE Initialize $T^* = \infty$, $\eta^* = \eta_{\min}$.
\FOR{$m = 0$ to $M-1$}
    \STATE $\eta = \eta_{\min} + m\cdot\Delta\eta$.
    \STATE Compute $J(\eta)$ via \eqref{eq10} and $Q(\eta)$ via \eqref{eq21}.
    \STATE Run Algorithm~\ref{alg:delay_bisection} to obtain $T^*(\eta)$.
    \IF{$T^*(\eta) < T^*$}
        \STATE $T^* = T^*(\eta)$, $\eta^* = \eta$.
    \ENDIF
\ENDFOR
\RETURN $\eta^*$, $T^*$, and $\{b_k, f_k, p_k, t_k\}_{k\in\tilde{\mathcal{K}}}$.
\end{algorithmic}
\end{algorithm}
\subsection{Complexity Analysis}

The computational complexity of Algorithm~\ref{alg:overall_simple} is determined by the hierarchical structure of nested loops. Specifically, the outermost layer involves a uniform sampling over the local accuracy parameter $\eta$, which is discretized into $M$ points. For each fixed $\eta$, the optimal delay $T$ is found via a bisection search over the interval $[T_{\min}, T_{\max}]$. Given a convergence tolerance $\epsilon_T$, this bisection search requires $\mathcal{O}(\log(1/\epsilon_T))$ iterations. Within each iteration of the delay search, the feasibility of all $K-1$ remaining clients must be verified. For each client $k$, the feasibility test involves a bisection search over the bandwidth $b_k$ to find the minimum required bandwidth $b_k^{\min}$, which exhibits a logarithmic convergence rate of $\mathcal{O}(\log(1/\epsilon_b))$. Furthermore, nested within this bandwidth search, the optimal transmission time $t_k$ is determined via a golden-section search, adding another complexity factor of $\mathcal{O}(\log(1/\epsilon_t))$. Aggregating these factors, the overall computational complexity of the proposed algorithm is 
\begin{equation}
    \mathcal{O} \left( M K \log\left(\frac{1}{\epsilon_T}\right) \log\left(\frac{1}{\epsilon_b}\right) \log\left(\frac{1}{\epsilon_t}\right) \right).
\end{equation}
This polynomial-logarithmic scaling indicates that the algorithm remains feasible for practical implementations in edge networks.

\section{Simulation Results}\label{sec.sr}
This section evaluates the performance of the proposed algorithm through numerical simulations. We consider a single-cell wireless network where the BS is positioned at the center of a $500 \times 500$ square meter area, serving $K$ user devices that are uniformly distributed within this region.

The simulation parameters are carefully configured to reflect practical system conditions. We utilize the publicly available blog feedback dataset~\cite{59}, consisting of $60,000$ samples, for the learning task of predicting comment counts. Each data sample is characterized by a feature dimension of $281$. For the local computation model, the number of CPU cycles required per bit, denoted as $C_k$, is randomly generated within the range $[1, 3] \times 10^4$ cycles/bit, and the effective capacitance parameter is set to $\kappa = 10^{-28}$. We adopt a step size $\delta = 0.1$ and a regularization parameter $\xi = 0.1$, with a target global accuracy of $\epsilon_0 = 10^{-3}$. Each user is allocated a local dataset of $D_k = 500$ samples, which are randomly selected from the dataset with equal probability.

The wireless communication environment is modeled with a path loss of $128.1 + 37.6 \log_{10}(d)$ dB (where $d$ is the distance in km) and log-normal shadowing with a standard deviation of $8$ dB. The background noise power spectral density (PSD) is $N_0 = -174$ dBm/Hz. Unless otherwise stated, the system bandwidth is $B = 20$ MHz, the number of users is $K = 21$, the transmit data size $s = 28.1$ kbits, the maximum transmit power is $p_k^{\max} = 20$ dBm, the maximum CPU frequency is $f_k^{\max} = 2.5$ GHz, and the maximum energy budget is $E_k^{\max} = 80$ J. All results are obtained by averaging over $1000$ independent Monte Carlo simulation runs. A comprehensive summary of the simulation parameters is provided in Table \ref{tab:sim_parameters}.
\begin{table}[!t]
\footnotesize
\renewcommand{\arraystretch}{1.2}
\caption{Simulation Parameters}
\label{tab:sim_parameters}
\centering
\setlength{\tabcolsep}{2pt}
\begin{tabular}{p{2.5cm} p{1.3cm} p{2.4cm}} 
\toprule
\textbf{Parameter} & \textbf{Symbol} & \textbf{Value} \\
\midrule
Dataset & -- & BlogFeedback \\
Feature dimension & -- & 281 \\
Number of users & $K$ & 21 \\
Local dataset size & $D_k$ & 500 \\
System bandwidth & $B$ & 20 MHz \\
Path loss model & -- & $128.1+37.6\log_{10}(d)$ dB \\
Shadowing std. dev. & -- & 8 dB \\
Noise PSD & $N_0$ & $-174$ dBm/Hz \\
Transmit data size & $s$ & 28.1 kbits \\
Max. transmit power & $p_k^{\max}$ & 20 dBm \\
Max. CPU frequency & $f_k^{\max}$ & 2.5 GHz \\
Energy budget & $E_k^{\max}$ & 80 J \\
CPU cycles per bit & $C_k$ & $[1,3]\!\times\!10^4$ cycles/bit \\
Step size & $\delta$ & 0.1 \\
Regularization & $\xi$ & 0.1 \\
Global accuracy & $\epsilon_0$ & $10^{-3}$ \\
Capacitance & $\kappa$ & $10^{-28}$ \\
\bottomrule
\end{tabular}
\end{table}

We will refer to Algorithm~\ref{alg:overall_simple} as \textbf{Proposed-FUN} and compare it with the following benchmark schemes:
\begin{itemize}
    \item \textbf{EB-FUN (Equal Bandwidth):} A scheme where the total bandwidth is equally allocated among all users, while computation frequency and accuracy are optimized.
    \item \textbf{FA-FUN (Fixed Accuracy):} A scheme where all users are forced to use a fixed local accuracy $\eta$, optimizing only for bandwidth and frequency.
    \item \textbf{Retrain:} A baseline scheme that ignores the pre-trained model and retrains the global model from scratch.
\end{itemize}

\subsection{Impact of System Resources}

\begin{figure}[t]
    \centering
    \includegraphics[width=0.8\linewidth]{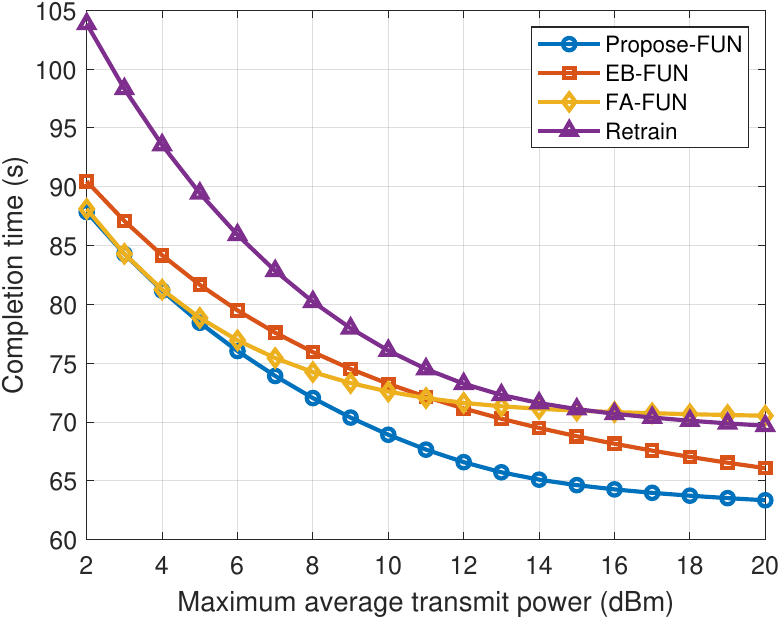}
    \caption{Completion time versus maximum average transmit power of each user.}
    \label{fig2}
\end{figure}

Fig.~\ref{fig2} illustrates the impact of the maximum average transmit power $p_k^{\max}$ on the completion time. As transmit power increases, the completion time for all schemes decreases, primarily due to the reduced communication latency resulting from higher data rates. `Propose-FUN’ consistently outperforms the baseline methods. Specifically, at a transmit power of 10 dBm, `Propose-FUN' reduces the completion time by approximately 10.7\% compared to `Retrain', 4.1\% compared to `EB-FUN', and 5.0\% compared to `FA-FUN'. This performance gain validates the effectiveness of our joint optimization strategy in balancing communication and computation resources. Furthermore, the performance gap tends to narrow at very high power levels, indicating that the task latency becomes dominated by computation time rather than transmission time in high-SNR regimes.

\begin{figure}[t]
    \centering
    \includegraphics[width=0.8\linewidth]{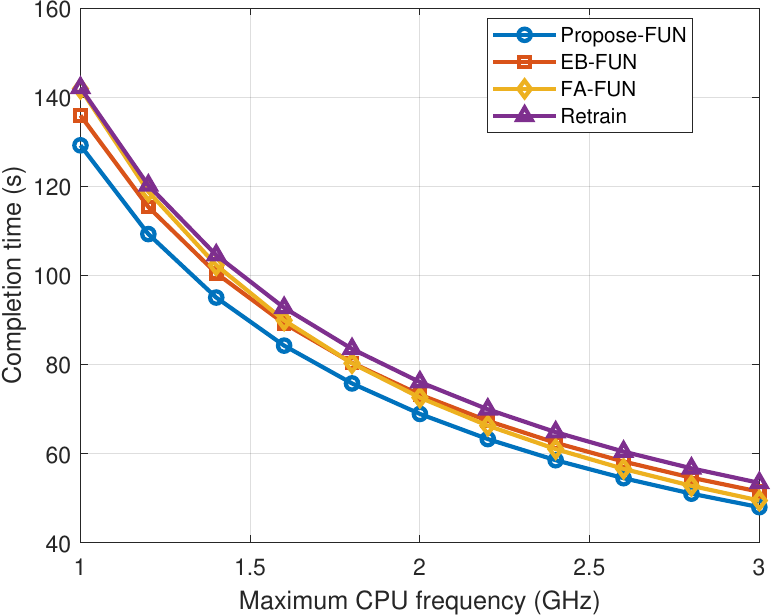}
    \caption{Completion time versus maximum CPU frequency for each user.}
    \label{fig3}
\end{figure}
\begin{figure}[t]
    \centering
    \includegraphics[width=0.8\linewidth]{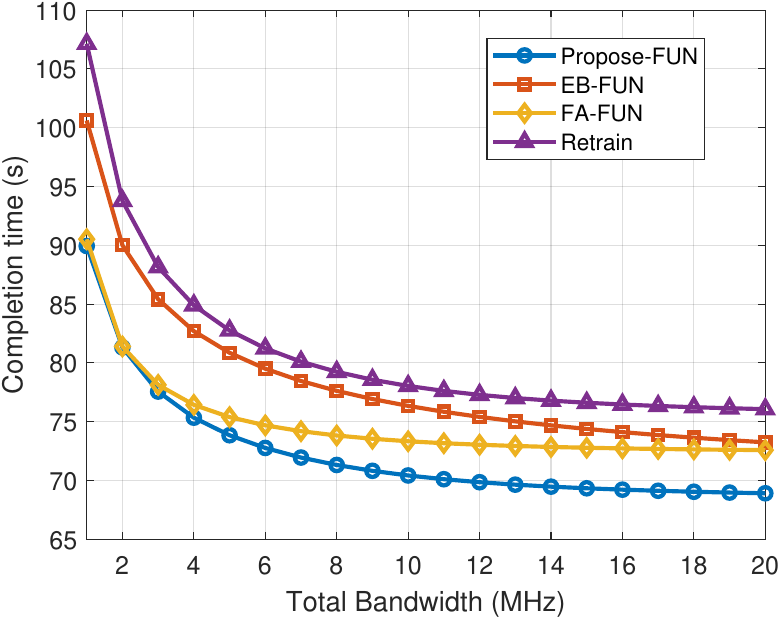}
    \caption{Completion time versus total bandwidth.}
    \label{fig4}
\end{figure}

Fig.~\ref{fig3} depicts the relationship between the maximum CPU frequency and the task completion time. As the computational capacity increases, the completion time for all schemes declines, attributed to the accelerated local processing speed. `Propose-FUN' demonstrates superior performance over the baselines. Specifically, at a CPU frequency of $2.0\text{ GHz}$, `Propose-FUN' reduces the completion time by approximately $10.5\%$ compared to `Retrain', $4.9\%$ compared to `EB-FUN', and $3.5\%$ compared to `FA-FUN'. These results highlight the robustness of our algorithm in optimizing computational resource allocation. Additionally, the slopes of the curves flatten as frequency increases, suggesting that the system performance becomes constrained by communication latency rather than computation power in this regime.

Fig.~\ref{fig4} demonstrates the effect of the total available bandwidth on the task completion time. The completion time for all schemes decreases as the bandwidth increases, which is primarily due to the enhanced data transmission rates that shorten the communication latency. `Propose-FUN' algorithm consistently achieves the lowest latency compared to the baselines. Specifically, at a total bandwidth of 10 MHz, `Propose-FUN' reduces the completion time by approximately 10.2\% compared to `Retrain', 7.2\% compared to `EB-FUN', and 4.1\% compared to `FA-FUN'. This highlights the robustness of our approach in leveraging spectrum resources efficiently. Furthermore, the performance curves tend to saturate when the bandwidth exceeds 12 MHz, indicating that the system performance becomes computation-limited rather than communication-limited in wideband scenarios.
\begin{figure}[t]
    \centering
    \includegraphics[width=0.8\linewidth]{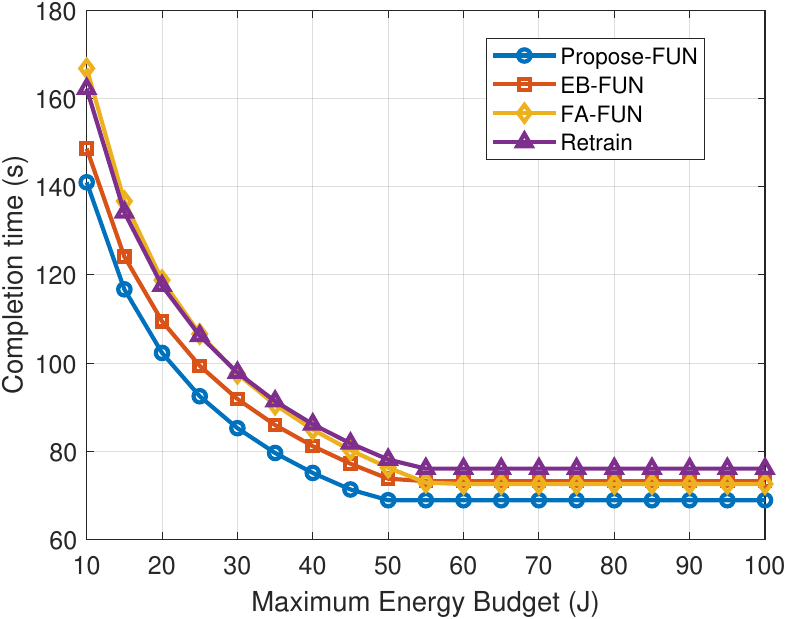}
    \caption{Completion time versus maximum energy budget of each user.}
    \label{fig5}
\end{figure}

Fig.~\ref{fig5} illustrates the impact of the maximum energy budget on task completion time. As the energy constraint is relaxed, the completion time for all schemes decreases significantly before stabilizing, as more energy allows for higher transmission power and computational frequency to accelerate task execution. `Propose-FUN' algorithm consistently achieves the lowest latency. Specifically, at an energy budget of 40 J, `Propose-FUN' reduces the completion time by approximately 8.9\% compared to `Retrain', 4.7\% compared to `EB-FUN', and 6.8\% compared to `FA-FUN'. Furthermore, the performance curves saturate when the energy budget exceeds 60 J, indicating that the latency becomes constrained by other factors, such as channel conditions or hardware limits, rather than energy availability.

\subsection{Impact of Network Scale and Data Size}
\begin{figure}[t]
    \centering
    \includegraphics[width=0.8\linewidth]{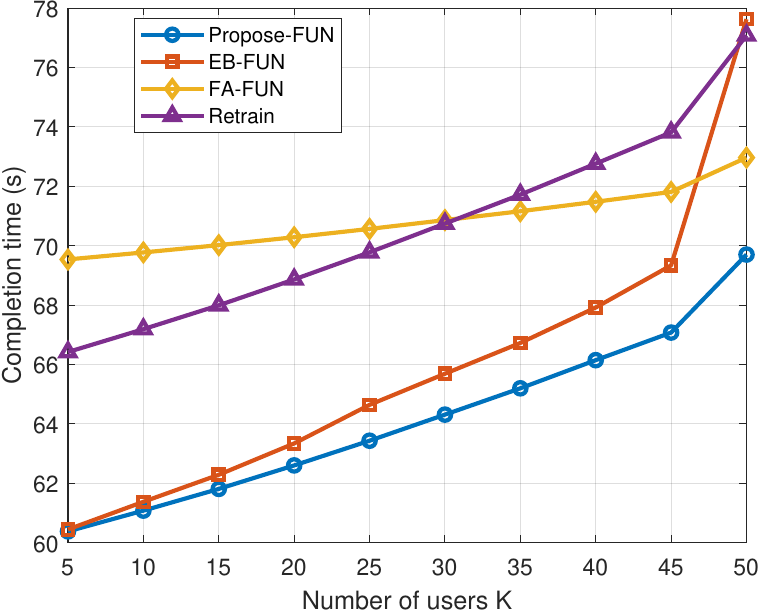}
    \caption{Completion time versus the number of users K.}
    \label{fig6}
\end{figure}
\begin{figure}[t]
    \centering
    \includegraphics[width=0.8\linewidth]{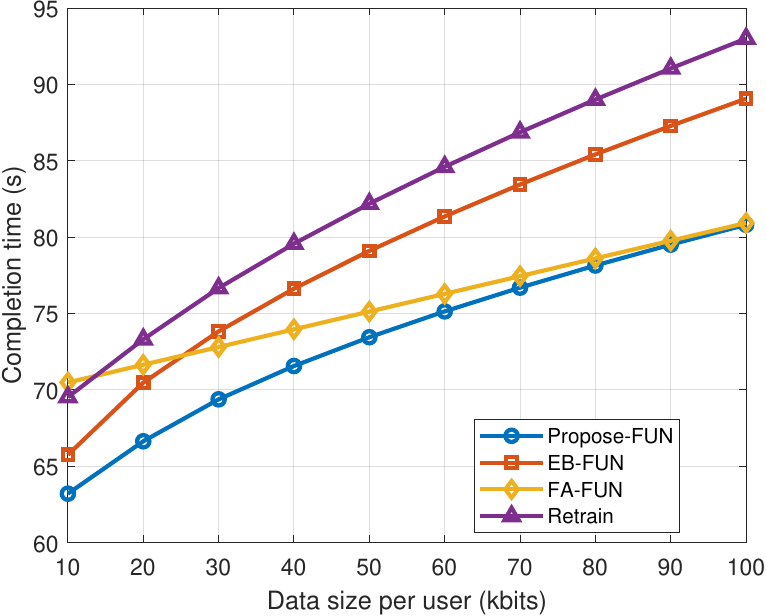}
    \caption{Completion time versus data size per user.}
    \label{fig7}
\end{figure}

Fig.~\ref{fig6} investigates the scalability of the proposed scheme by varying the number of users $K$. As the number of users increases, the completion time for all schemes rises due to the heavier computational load and the limited radio resources allocated per user. However, `Propose-FUN' consistently maintains the lowest latency compared to the benchmarks. Specifically, at $K=40$ , `Propose-FUN' reduces the completion time by approximately 8.9\% compared to `Retrain', 2.9\% compared to `EB-FUN', and 7.0\% compared to `FA-FUN'. Furthermore, it is worth noting that the performance of Retrain degrades sharply when $K>45$ , whereas `Propose-FUN' exhibits a more moderate growth rate, demonstrating its robustness and effectiveness in handling dense user scenarios.

Fig.~\ref{fig7} depicts the impact of the data size per user on the task completion time. As the data size increases, the completion time for all schemes exhibits a linear upward trend, primarily due to the increased computational load and the longer transmission time required for larger data payloads. `Propose-FUN' consistently achieves the lowest latency across all data sizes. Specifically, at a data size of 60 kbits, `Propose-FUN' reduces the completion time by approximately 10.6\% compared to `Retrain', 6.7\% compared to `EB-FUN', and 1.3\% compared to `FA-FUN'. 

\subsection{Impact of Local Accuracy and Trade-offs}

\begin{figure}[t]
    \centering
    \includegraphics[width=0.8\linewidth]{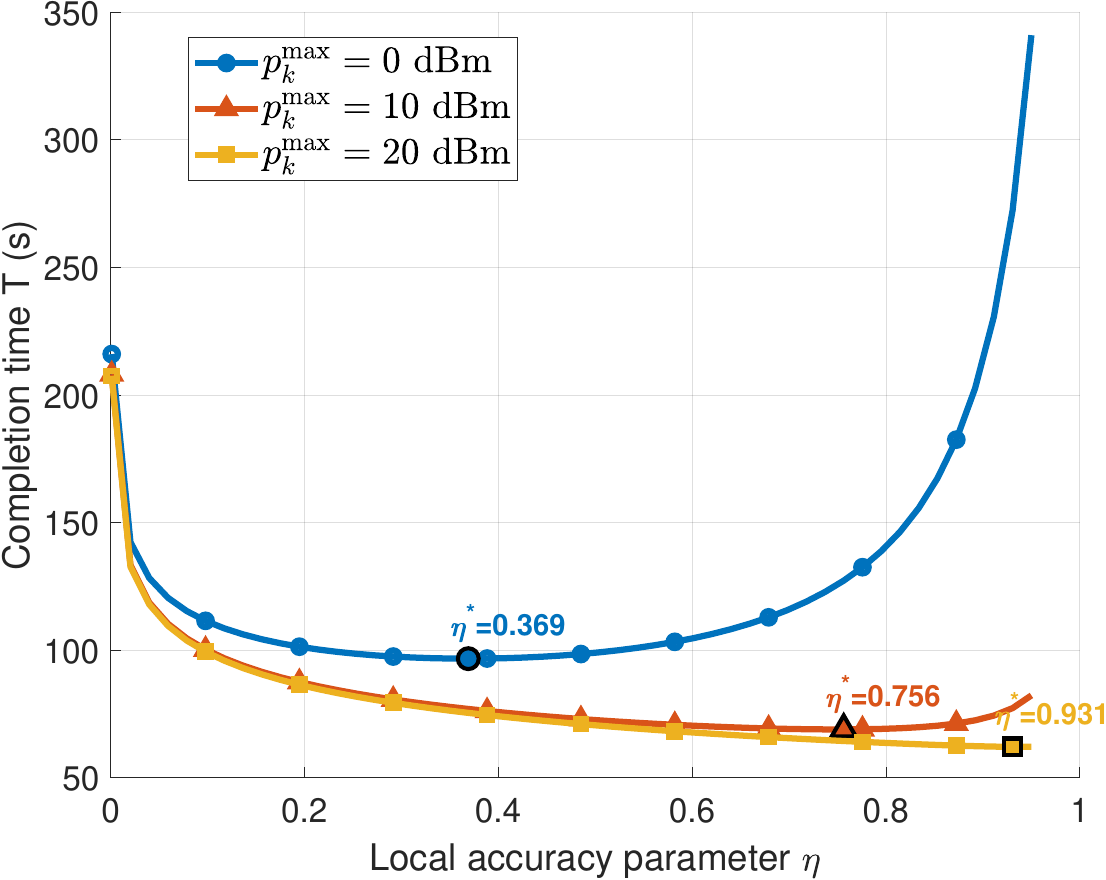}
    \caption{Completion time versus maximum average transmit power of each user.}
    \label{fig8}
\end{figure}

Fig.~\ref{fig8} illustrates the trade-off between the local accuracy $\eta$ and the task completion time under different maximum transmit power constraints. The curves exhibit a convex shape, indicating that there exists an optimal $\eta^*$ that minimizes the completion time. Specifically, for low power constraints, the optimal $\eta^*$ is small, as the slow transmission rate makes excessive local computation inefficient. Conversely, as the transmit power increases to 20 dBm , the optimal $\eta^*$ shifts to a higher value, and the minimum completion time decreases significantly. This demonstrates that with faster communication capabilities, the system benefits from more local computation to reduce the number of communication rounds.

\section{Conclusion}\label{sec.c}
In this paper, we have investigated the problem of delay minimization for FUN. We have derived the delay and energy consumption models for FUN based on the convergence rate and robust transmission constraints. With these models, we have formulated a joint optimization problem to minimize the maximum per-client delay under energy and bandwidth constraints. To solve this non-convex problem, we have proposed an iterative algorithm with low complexity, which decomposes the problem via a uniform scan over the local accuracy parameter and employs nested bisection and golden-section searches. Simulation results have shown that the proposed scheme significantly outperforms conventional schemes in terms of completion time, especially in scenarios with limited energy budgets and imperfect channel state information. In future work, we plan to extend this framework to scenarios with multiple simultaneous unlearning requests and dynamic time-varying channel conditions.

\begin{appendices}
\numberwithin{equation}{section}
\section{PROOF OF LEMMA 1}
Based on \eqref{eq10} and Lemma 5 in \cite{56}, the following conditions hold 
\begin{align}\label{eqA1}
    \frac{1}{L}&\left\|\nabla F_k(\boldsymbol{\widetilde{w}}^{(j)} + \boldsymbol{h}^{(j)}) - \nabla F_k(\boldsymbol{\widetilde{w}}^{(j)})\right\|^2 \nonumber\\ &\leq (\nabla F_k(\boldsymbol{\widetilde{w}}^{(j)} + \boldsymbol{h}^{(j)}) - \nabla F_k(\boldsymbol{\widetilde{w}}^{(j)}))^T\boldsymbol{h}^{(j)} \nonumber\\ &\leq \frac{1}{\gamma} ||\nabla F_k(\boldsymbol{\widetilde{w}}^{(j)} + \boldsymbol{h}^{(j)}) - \nabla F_k(\boldsymbol{\widetilde{w}}^{(j)})||^2,
\end{align}
and
\begin{align}\label{eqA2}
   ||\nabla F(w)||^2 \geq \gamma \left(F(\boldsymbol{w})-F(\boldsymbol{w}^*)\right).
\end{align}

For the optimal solution of local problem, the first-order derivative condition always holds,i.e.,
\begin{align}\label{eqA3}
   \nabla G_k&(\widetilde{\boldsymbol w}^{(j)},\hat{\boldsymbol h}_k^{(j)*})\nonumber\\
   &=\nabla F_k(\boldsymbol{\widetilde{w}}^{(j)}+\boldsymbol{\hat{h}}_k^{(j)*})-\nabla F_k(\boldsymbol{\widetilde{w}}^{(j)})+\xi\nabla F(\boldsymbol{\widetilde{w}}^{(j)})\nonumber\\
   &=\boldsymbol{0}.
\end{align}
With the above inequalities and equalities, we have
\begin{align}\label{eqA4}
    F&(\boldsymbol{\widetilde{w}}^{(j+1)})\nonumber\\ 
    &\overset{\eqref{eq7},\eqref{eq10}}{\leq} F(\boldsymbol{\widetilde{w}}^{(j)})+\frac{1}{K-1}\sum_{k \in \mathcal{\widetilde{K}}}\nabla F(\boldsymbol{\widetilde{w}}^{(j)})^T \boldsymbol{\widetilde{h}}_k^{(j)}\nonumber\\
    & \quad+\frac{L}{2(K-1)^2}\left\| \sum_{k \in \mathcal{\widetilde{K}}}\boldsymbol{\widetilde{h}}_k^{(j)}\right\|^2\nonumber\\
    &\overset{\eqref{eq6}}{=} F(\boldsymbol{\widetilde{w}}^{(j)})+\frac{1}{K-1}\sum_{k \in \mathcal{\widetilde{K}}}\frac{\left\|\boldsymbol{h}_k^{(j)}\right\|}{\left\|\boldsymbol{\hat{h}}_k^{(j)}\right\|}\nabla F(\boldsymbol{\widetilde{w}}^{(j)})^T \boldsymbol{\hat{h}}_k^{(j)}\nonumber\\
    & \quad+\frac{L}{2(K-1)^2}\left\| \sum_{k \in \mathcal{\widetilde{K}}}\boldsymbol{\widetilde{h}}_k^{(j)}\right\|^2\nonumber\\
    &\overset{\eqref{eq3}}{=} F(\boldsymbol{\widetilde{w}}^{(j)})+\frac{1}{(K-1)\xi}\sum_{k \in \mathcal{\widetilde{K}}}\frac{\left\|\boldsymbol{h}_k^{(j)}\right\|}{\left\|\boldsymbol{\hat{h}}_k^{(j)}\right\|}\Big[G_k(\boldsymbol{\widetilde{w}}^{(j)},\boldsymbol{\hat{h}}_k^{(j)})\nonumber\\ 
    & \quad +\nabla F_k(\boldsymbol{\widetilde{w}}^{(j)})\boldsymbol{\hat{h}}_k^{(j)}-F_k(\boldsymbol{\widetilde{w}}^{(j)}+\boldsymbol{\hat{h}}_k^{(j)})\Big]\nonumber\\ 
    &\quad +\frac{L}{2(K-1)^2}\left\| \sum_{k \in \mathcal{\widetilde{K}}}\boldsymbol{\widetilde{h}}_k^{(j)}\right\|^2\nonumber\\
    &\overset{\eqref{eq10}}{\leq} F(\boldsymbol{\widetilde{w}}^{(j)})+\frac{1}{(K-1)\xi}\sum_{k \in \mathcal{\widetilde{K}}}\frac{\left\|\boldsymbol{h}_k^{(j)}\right\|}{\left\|\boldsymbol{\hat{h}}_k^{(j)}\right\|}\Big[ G_k(\boldsymbol{\widetilde{w}}^{(j)},\boldsymbol{\hat{h}}_k^{(j)})\nonumber\\ 
    &\quad-F_k(\boldsymbol{\widetilde{w}}^{(j)})-\frac{\gamma}{2}\left\|\boldsymbol{\hat{h}}_k^{(j)}\right\|^2\Big]+\frac{L}{2(K-1)^2}\left\| \sum_{k \in \mathcal{\widetilde{K}}}\boldsymbol{\widetilde{h}}_k^{(j)}\right\|^2.
\end{align}
According to the triangle inequality and mean inequality, we have
\begin{align}\label{eqA5}
   \left\|\frac{1}{K-1}\sum_{k \in \mathcal{\widetilde{K}}}\boldsymbol{\widetilde{h}}_k^{(j)}\right\|^2 &\leq \left(\frac{1}{K-1}\sum_{k \in \mathcal{\widetilde{K}}}\left\|\boldsymbol{\widetilde{h}}_k^{(j)}\right\|\right)^2\nonumber\\
   &\leq \frac{1}{K-1}\sum_{k \in \mathcal{\widetilde{K}}}\left\|\boldsymbol{\widetilde{h}}_k^{(j)}\right\|^2
\end{align}
By the properties of norms, we can infer that
\begin{align}\label{eqA6}
\left\|\boldsymbol{\widetilde{h}}_k^{(j)}\right\|^2 = \left\|\boldsymbol{h}_k^{(j)}\right\|^2\left\|\frac{ \boldsymbol{\hat{h}}_k^{(j)}}{ \left\|\boldsymbol{\hat{h}}_k^{(j)}\right\|}\right\|^2 =  \left\|\boldsymbol{h}_k^{(j)}\right\|^2.
\end{align}
Combining \eqref{eqA4} and \eqref{eqA5} yields
\begin{align}\label{eqA7}
   F&(\boldsymbol{\widetilde{w}}^{(j+1)})\nonumber\\ 
   &\leq F(\boldsymbol{\widetilde{w}}^{(j)})+\frac{1}{(K-1)\xi}\sum_{k \in \mathcal{\widetilde{K}}}\frac{\left\|\boldsymbol{h}_k^{(j)}\right\|}{\left\|\boldsymbol{\hat{h}}_k^{(j)}\right\|}\Big[ G_k(\boldsymbol{\widetilde{w}}^{(j)},\boldsymbol{\hat{h}}_k^{(j)})\nonumber\\ 
    &\quad-F_k(\boldsymbol{\widetilde{w}}^{(j)})-\frac{\gamma}{2}\left\|\boldsymbol{\hat{h}}_k^{(j)}\right\|^2\Big]+\frac{L}{2(K-1)} \sum_{k \in \mathcal{\widetilde{K}}}\left\|\boldsymbol{h}_k^{(j)}\right\|^2.
\end{align}
Applying \eqref{eq9} to \eqref{eqA7}, we can obtain
\begin{align}\label{eqA8}
    F&(\boldsymbol{\widetilde{w}}^{(j+1)})\nonumber\\ 
    &\leq F(\boldsymbol{\widetilde{w}}^{(j)})+\frac{\beta}{(K-1)\xi}\sum_{k \in \mathcal{\widetilde{K}}}\Big[ G_k(\boldsymbol{\widetilde{w}}^{(j)},\boldsymbol{\hat{h}}_k^{(j)})\nonumber\\ 
    &\quad-F_k(\boldsymbol{\widetilde{w}}^{(j)})-\frac{\gamma-L\xi\beta}{2}\left\|\boldsymbol{\hat{h}}_k^{(j)}\right\|^2\Big]\nonumber\\
    &\overset{\eqref{eq3}}{=} F(\boldsymbol{\widetilde{w}}^{(j)})+\frac{\beta}{(K-1)\xi}\sum_{k \in \mathcal{\widetilde{K}}}\Big[ G_k(\boldsymbol{\widetilde{w}}^{(j)},\boldsymbol{\hat{h}}_k^{(j)})\nonumber\\ 
    &\quad-G_k(\boldsymbol{\widetilde{w}}^{(j)},\boldsymbol{\hat{h}}_k^{(j)*}) -(G_k(\boldsymbol{\widetilde{w}}^{(j)},\boldsymbol{0})-G_k(\boldsymbol{\widetilde{w}}^{(j)},\boldsymbol{\hat{h}}_k^{(j)*}))\nonumber\\
    &\quad-\frac{\gamma-L\xi\beta}{2}\left\|\boldsymbol{\hat{h}}_k^{(j)}\right\|^2\Big]\nonumber\\
    &\overset{\eqref{eq5}}{\leq} F(\boldsymbol{\widetilde{w}}^{(j)})-\frac{\beta}{(K-1)\xi}\sum_{k \in \mathcal{\widetilde{K}}}\Big[ (1-\eta) (G_k(\boldsymbol{\widetilde{w}}^{(j)},\boldsymbol{0})\nonumber\\
    &\quad-G_k(\boldsymbol{\widetilde{w}}^{(j)},\boldsymbol{\hat{h}}_k^{(j)*}))+\frac{\gamma-L\xi\beta}{2}\left\|\boldsymbol{\hat{h}}_k^{(j)}\right\|^2\Big]\nonumber\\
    & \overset{\eqref{eq3}}{=} F(\boldsymbol{\widetilde{w}}^{(j)})-\frac{\beta}{(K-1)\xi}\sum_{k \in \mathcal{\widetilde{K}}}\Big[ (1-\eta) \big( F_k(\boldsymbol{\widetilde{w}}^{(j)})\nonumber\\
    &\quad-F_k(\boldsymbol{\widetilde{w}}^{(j)}+\boldsymbol{\hat{h}}_k^{(j)*})+(\nabla F_k(\boldsymbol{
    \widetilde{w}}^{(j)})\nonumber\\
    &\quad-\xi\nabla F(\boldsymbol{\widetilde{w}}^{(j)}))^T\boldsymbol{\hat{h}}_k^{(j)*} \big)+\frac{\gamma-L\xi\beta}{2}\left\|\boldsymbol{\hat{h}}_k^{(j)}\right\|^2\Big]\nonumber\\
    & \overset{\eqref{eqA3}}{=} F(\boldsymbol{\widetilde{w}}^{(j)})-\frac{\beta}{(K-1)\xi}\sum_{k \in \mathcal{\widetilde{K}}}\Big[ (1-\eta) \big( F_k(\boldsymbol{\widetilde{w}}^{(j)})\nonumber\\
    &\quad-F_k(\boldsymbol{\widetilde{w}}^{(j)}+\boldsymbol{\hat{h}}_k^{(j)*})+\nabla F_k(\boldsymbol{\widetilde{w}}^{(j)} +\boldsymbol{\hat{h}}_k^{(j)*})^T\boldsymbol{\hat{h}}_k^{(j)*} \big)\nonumber\\
    &\quad+\frac{\gamma-L\xi\beta}{2}\left\|\boldsymbol{\hat{h}}_k^{(j)}\right\|^2\Big].
\end{align}

Base on \eqref{eq10}, we can obtain
\begin{align}\label{eqA9}
   F_k(\boldsymbol{\widetilde{w}}^{(j)})\geq &F_k(\boldsymbol{\widetilde{w}}^{(j)}+\boldsymbol{\hat{h}}_k^{(j)*})\nonumber\\
   &+\nabla F_k(\boldsymbol{\widetilde{w}}^{(j)}+\boldsymbol{\hat{h}}_k^{(j)*})^T(-\boldsymbol{\hat{h}}_k^{(j)*})+\frac{\gamma}{2}\left\|\boldsymbol{\hat{h}}_k^{(j)*}\right\|^2.
\end{align}
Applying \eqref{eqA9} to \eqref{eqA8}, we have
\begin{align}\label{eqA10}
   F_k(\boldsymbol{\widetilde{w}}^{(j+1)})&\leq F(\boldsymbol{\widetilde{w}}^{(j)})-\frac{\beta}{(K-1)\xi}\sum_{k=1,k\neq k_u}^K\Big[\frac{(1-\eta)\gamma}{2}\nonumber\\
   &\quad\times\left\|\boldsymbol{\hat{h}}_k^{(j)*}\right\|^2+\frac{\gamma-L\xi\beta}{2}\left\|\boldsymbol{\hat{h}}_k^{(j)}\right\|^2\Big].
\end{align}

Base on \eqref{eq3} and \eqref{eq10}, $F_k(\cdot)$ is $L$-Lipschitz continuous and $\gamma$-strongly convex, and hence $G_k(\widetilde{\boldsymbol w}^{(j)},\boldsymbol {\hat{h}}_k)$ is also $L$-Lipschitz continuous and $\gamma$-strongly covex with respect to $\boldsymbol {\hat{h}}_k$. 

By the definition of $L$-Lipschitz continuous and $\gamma$-strongly covex, for any $\boldsymbol {\hat{h}}_k$, we have
\begin{align}\label{eqA11}
G_k&(\widetilde{\boldsymbol w}^{(j)},\boldsymbol {0})-G_k(\widetilde{\boldsymbol w}^{(j)},\hat{\boldsymbol h}_k^{(j)*})\nonumber\\
&\leq \nabla G_k(\widetilde{\boldsymbol w}^{(j)},\hat{\boldsymbol h}_k^{(j)*})^T(\boldsymbol {0}-\hat{\boldsymbol h}_k^{(j)*})+\frac{L}{2}\left\|\boldsymbol {0}-\hat{\boldsymbol h}_k^{(j)*}\right\|^2\nonumber\\
&\overset{\eqref{eqA3}}{=}\frac{L}{2}\left\|\hat{\boldsymbol h}_k^{(j)*}\right\|^2,
\end{align}
and
\begin{align}\label{eqA12}
G_k&(\widetilde{\boldsymbol w}^{(j)},\boldsymbol {\hat{h}}_k)-G_k(\widetilde{\boldsymbol w}^{(j)},\hat{\boldsymbol h}_k^{(j)*})\nonumber\\
&\geq \nabla G_k(\widetilde{\boldsymbol w}^{(j)},\hat{\boldsymbol h}_k^{(j)*})^T(\boldsymbol {\hat{h}}_k-\hat{\boldsymbol h}_k^{(j)*})+\frac{\gamma}{2}\left\|\boldsymbol {\hat{h}}_k-\hat{\boldsymbol h}_k^{(j)*}\right\|^2\nonumber\\
&\overset{\eqref{eqA3}}{=}\frac{\gamma}{2}\left\|\boldsymbol {\hat{h}}_k-\hat{\boldsymbol h}_k^{(j)*}\right\|^2.
\end{align}
Base on \eqref{eq5} , we can obtain
\begin{align}\label{eqA13}
\frac{\gamma}{2}\left\|\hat{\boldsymbol h}_k^{(j)}-\hat{\boldsymbol h}_k^{(j)*}\right\|^2\le\frac{\eta L}{2}\left\|\hat{\boldsymbol h}_k^{(j)*}\right\|^2.
\end{align}
Formula \eqref{eqA13} can be simplified as
\begin{align}\label{eqA14}
\left\|\hat{\boldsymbol h}_k^{(j)}-\hat{\boldsymbol h}_k^{(j)*}\right\|\le\sqrt{\frac{\eta L}{\gamma}}\left\|\hat{\boldsymbol h}_k^{(j)*}\right\|.
\end{align}
By the reverse triangle inequality, we have
\begin{align}\label{eqA15}
\left\|\hat{\boldsymbol h}_k^{(j)}\right\|\ge&\left\|\hat{\boldsymbol h}_k^{(j)*}\right\|-\left\|\hat{\boldsymbol h}_k^{(j)}-\hat{\boldsymbol h}_k^{(j)*}\right\|\nonumber \\
\overset{\eqref{eqA14}}{\ge}&\left(1-\sqrt{\frac{\eta L}{\gamma}}\right)\left\|\hat{\boldsymbol h}_k^{(j)*}\right\|.
\end{align}
From \eqref{eqA1}, the following relationship is obtained 
\begin{align}\label{eqA16}
\left\|\boldsymbol{\hat{h}}_k^{(j)*}\right\|^2\geq\frac{1}{L^2}\left\|\nabla F_k(\boldsymbol{\widetilde{w}}^{(j)}+\boldsymbol{\hat{h}}_k^{(j)*})-\nabla F_k(\boldsymbol{\widetilde{w}}^{j})\right\|^2.
\end{align}
For the constant parameter $\xi$, we choose 
\begin{align}\label{eqA17}
\gamma-L\xi\beta>0.
\end{align}

According to \eqref{eqA10}, \eqref{eqA14} and \eqref{eqA17}, we can obtain
\begin{align}\label{eqA18}
F_k&(\boldsymbol{\widetilde{w}}^{(j+1)})\nonumber\\&\leq F(\boldsymbol{\widetilde{w}}^{(j)})-\frac{\beta}{(K-1)\xi}\nonumber\\
&\quad\times\frac{(\gamma-L\xi\beta)(\sqrt{\gamma}-\sqrt{\eta L})^2+(1-\eta)\gamma^2}{2\gamma}\nonumber\\
&\quad\times\sum_{k \in \mathcal{\widetilde{K}}}\left\|\boldsymbol{\hat{h}}_k^{(j)*}\right\|^2\nonumber\\
&\overset{\eqref{eqA16}}{\leq} F(\boldsymbol{\widetilde{w}}^{(j)})\nonumber\\
&\quad-\frac{\beta\big[(\gamma-L\xi\beta)(\sqrt{\gamma}-\sqrt{\eta L})^2+(1-\eta)\gamma^2\big]}{2\gamma\xi L^2(K-1)}\nonumber\\
&\quad\times\sum_{k \in \mathcal{\widetilde{K}}}\left\|\nabla F_k(\boldsymbol{\widetilde{w}}^{(j)}+\boldsymbol{\hat{h}}_k^{(j)*})-\nabla F_k(\boldsymbol{\widetilde{w}}^{j})\right\|^2\nonumber\\
&\overset{\eqref{eqA3}}{=}F(\boldsymbol{\widetilde{w}}^{(j)})\nonumber\\
&\quad-\frac{\xi\beta\big[(\gamma-L\xi\beta)(\sqrt{\gamma}-\sqrt{\eta L})^2+(1-\eta)\gamma^2\big]}{2\gamma L^2}\nonumber\\
&\quad\times\left\|\nabla F(\boldsymbol{\widetilde{w}}^{(j)})\right\|^2\nonumber\\
&\overset{\eqref{eqA2}}{\leq} F(\boldsymbol{\widetilde{w}}^{(j)})\nonumber\\
&\quad-\frac{\xi\beta\big[(\gamma-L\xi\beta)(\sqrt{\gamma}-\sqrt{\eta L})^2+(1-\eta)\gamma^2\big]}{2 L^2}\nonumber\\
&\quad\times (F(\boldsymbol{\widetilde{w}}^{(j)})-F(\boldsymbol{\widetilde{w}}^*)).
\end{align}

Base on \eqref{eqA18}, we get
\begin{align}\label{eqA19}
&F_k(\boldsymbol{\widetilde{w}}^{(j+1)})-F(\boldsymbol{\widetilde{w}}^*)\nonumber\\
&\leq \left(1-\frac{\xi\beta\big[(\gamma-L\xi\beta)(\sqrt{\gamma}-\sqrt{\eta L})^2+(1-\eta)\gamma^2\big]}{2 L^2}\right)\nonumber\\
&\quad\times (F(\boldsymbol{\widetilde{w}}^{(j)})-F(\boldsymbol{\widetilde{w}}^*))\nonumber\\
&\leq \left(1-\frac{\xi\beta\big[(\gamma-L\xi\beta)(\sqrt{\gamma}-\sqrt{\eta L})^2+(1-\eta)\gamma^2\big]}{2 L^2}\right)^{j+1}\nonumber\\
&\quad\times (F(\boldsymbol{\widetilde{w}}^{(0)})-F(\boldsymbol{\widetilde{w}}^*))\nonumber\\
&\leq \mathrm{exp}{\left(-(j+1)\frac{\xi\beta\big[(\gamma-L\xi\beta)(\sqrt{\gamma}-\sqrt{\eta L})^2+(1-\eta)\gamma^2\big]}{2 L^2}\right)}\nonumber\\
&\quad\times (F(\boldsymbol{\widetilde{w}}^{(0)})-F(\boldsymbol{\widetilde{w}}^*)).
\end{align}

To ensure that $F(\boldsymbol{\tilde{w}}^{(j+1)})-F(\boldsymbol{\tilde{w}}^*)\leq \epsilon_0(F(\boldsymbol{\tilde{w}}^{(0)})-F(\boldsymbol{\tilde{w}}^*))$, we have

\begin{align}\label{eqA20}
\mathrm{exp}&{\left(-(j+1)\frac{\xi\beta\big[(\gamma-L\xi\beta)(\sqrt{\gamma}-\sqrt{\eta L})^2+(1-\eta)\gamma^2\big]}{2 L^2}\right)}\nonumber\\
&\leq \epsilon_0, 
\end{align}
which can be simplified to \eqref{eq11}.

\section{PROOF OF LEMMA 3}
To demonstrate the monotonicity of feasibility with respect to the allocated bandwidth $b_k$, we analyze the sensitivity of the total energy consumption $E_k^\text{total}(t_k)$ to changes in $b_k$.

Recall from Section III-B that the total energy consumption for user $k\in \widetilde{\mathcal{K}}$ is the sum of computation and communication energy, given by
\begin{equation}
    E_k^\text{total}(t_k) = \underbrace{\kappa \left(f_k^{\min}(t_k)\right)^2 Q(\eta) A_k}_{\text{Computation Energy}} + \underbrace{J(\eta) p_k^{\min}(t_k) t_k}_{\text{Communication Energy}},
\end{equation}
where $f_k^{\min}(t_k)$ and $p_k^{\min}(t_k)$ are obtained from \eqref{eq22} and \eqref{eq23} respectively. Note that for a fixed $t_k$, the computation energy is independent of the bandwidth $b_k$.

Thus, the communication energy $E_k^\text{comm}$ can be expressed as
\begin{equation}
    E_k^\text{comm}(t_k, b_k) = \frac{J(\eta) N_0 b_k t_k}{(\left| \hat{g}_k\right|-\epsilon_k)^2} \left( 2^{\frac{s}{t_k b_k}} - 1 \right).
\end{equation}

Let us analyze the monotonicity of $E_k^\text{comm}$ with respect to $b_k$ for a fixed $t_k$. Let $u = t_k b_k$. The term governing the energy behavior is proportional to $b_k \left( 2^{\frac{s}{t_k b_k}} - 1 \right)$. It is a standard result in communication theory that for a fixed transmission time and data payload, the required transmission energy is a strictly decreasing function of the bandwidth. Mathematically, we have
\begin{equation}
    \frac{\partial E_k^\text{comm}(t_k, b_k)}{\partial b_k} < 0.
\end{equation}

Since the computation energy is constant with respect to $b_k$ and the communication energy decreases as $b_k$ increases, the total energy $E_k^\text{total}(t_k)$ is monotonically decreasing with respect to $b_k$.

Therefore, if user $k$ is feasible at bandwidth $b_k$ (i.e., there exists a $t_k$ such that $E_k^\text{total}(t_k, b_k) \le E_k^{\max}$), then for any $b_k' \ge b_k$, the energy consumption at the same $t_k$ satisfies:
\begin{equation}
    E_k^\text{total}(t_k)(t_k, b_k') < E_k^\text{total}(t_k, b_k) \le E_k^{\max}.
\end{equation}
This implies that the feasibility region is non-contracting with increased bandwidth.

\end{appendices}

\end{document}